\magnification=1200
\def\tr{{\rm tr}}
\def\f#1#2{{\textstyle{#1\over #2}}}
\def\next{\hfil\break\noindent}
\font\title=cmbx12

{\title 
\centerline{Constant mean curvature foliations in cosmological spacetimes}}

\vskip 1cm

\noindent
Alan D. Rendall
\next
Max-Planck-Institut f\"ur Gravitationsphysik
\next
Schlaatzweg 1
\next
14473 Potsdam
\next
Germany

\vskip 1cm\noindent
{\bf Abstract} 

\noindent
Foliations by constant mean curvature hypersurfaces provide
a possibility of defining a preferred time coordinate in general
relativity. In the following various conjectures are made about the
existence of foliations of this kind in spacetimes satisfying the strong 
energy condition and possessing compact Cauchy hypersurfaces. Recent
progress on proving these conjectures under supplementary assumptions
is reviewed. The method of proof used is explained and the prospects
for generalizing it discussed. The relations of these questions to 
cosmic censorship and the closed universe recollapse conjecture are
pointed out.

\vskip 1cm\noindent
{\bf 1. Conjectures on constant mean curvature hypersurfaces}

The purpose of this paper is to put forward some conjectures on the existence
of hypersurfaces of constant mean curvature in spatially compact spacetimes
and to present some recent results which show that these conjectures are
true in certain special cases. The spacetimes considered are globally 
hyperbolic solutions of the Einstein equations with vanishing cosmological
constant which possess a compact Cauchy hypersurface and satisfy the strong 
energy condition, i.e. the condition that $T_{\alpha\beta}u^\alpha u^\beta+
\f12 T^\alpha_\alpha\ge 0$ for any unit timelike vector $u^\alpha$. Following 
the terminology of Bartnik [1],
I refer to these as {\it cosmological spacetimes}. If $S$ is a spacelike
hypersurface in a spacetime $(M,g_{\alpha\beta})$, its induced metric and
second fundamental form will be denoted by $g_{ab}$ and $k_{ab}$ 
respectively. The mean curvature of $S$ is the trace $\tr k=g^{ab}k_{ab}$. 
The hypersurface $S$ is said to have constant mean curvature (CMC) if the 
function $\tr k$ on $S$ is constant. 

There are a number of well-known properties of CMC hypersurfaces in
cosmological spacetimes (see e.g. [1], [2]). The first concerns uniqueness:
in a cosmological spacetime there exists at most one compact hypersurface
with a given non-zero value of $\tr k$. In the case of a maximal hypersurface
($\tr k=0$) the statement is not quite so strong. In that case there is at
most one compact hypersurface with the given value of $\tr k$ unless the
spacetime is static with timelike Killing vector $t^\alpha$ and 
$R_{\alpha\beta}t^\alpha t^\beta=0$. For solutions of the Einstein equations
coupled to reasonable matter the latter situation is rare. For instance,
if the matter satisfies the non-negative pressures condition 
($T_{\alpha\beta}x^\alpha x^\beta\ge 0$ for any spacelike vector $x^\alpha$)
and the dominant energy condition, then a spacetime of this type is 
necessarily flat. The next property is that if there exists one compact
CMC hypersurface in a cosmological spacetime, there exists a foliation of a 
neighbourhood of that hypersurface by compact CMC hypersurfaces. The
mean curvature varies in a monotone way from slice to slice. 

These properties suggest one possible motivation for studying foliations by
CMC hypersurfaces. If such a foliation exists globally in a given spacetime
a function $t$ can be defined by the property that its value at each point 
of a given leaf of the foliation is equal to the mean curvature of that leaf.
This defines a scalar function on spacetime whose gradient is everywhere
timelike or zero. In fact it can be shown that where $t$ is non-zero its
gradient is timelike. Only on a maximal hypersurface can the gradient vanish
and even in that case it does not usually happen. Thus, leaving aside 
possible problems with exceptional maximal hypersurfaces, the CMC foliation
provides a global time coordinate which is invariantly defined by the geometry.
This is very useful if one wishes to consider the Einstein equation as an
evolution equation, for instance when investigating cosmic censorship [3].
It may also be of relevance for the problem of time in quantum gravity 
(cf. [4]). The exceptional maximal hypersurfaces, where $t$ is not a good
time coordinate have the properties that $k_{ab}=0$ and $R_{\alpha\beta}
n^\alpha n^\beta=0$, where $n^\alpha$ is the unit normal vector to the
hypersurfaces. These conditions are reminiscent of those for non-uniqueness
of maximal hypersurfaces mentioned earlier. As in that case, if the 
non-negative pressures and dominant energy conditions are satisfied then the
spacetime must be vacuum. Nevertheless there are many vacuum examples, e.g.
the Taub-NUT solution with $m=0$.

These general results do not immediately give any information about the 
existence
question, i.e. the question of whether a given cosmological spacetime
admits a CMC hypersurface. If a spacetime does admit a CMC hypersurface
one can ask which real numbers occur as the mean curvature of such a 
hypersurface in a given spacetime. Some light is thrown on the latter
question by a theorem on barrier hypersurfaces. Two compact hypersurfaces
$S_1$ and $S_2$ in a cosmological spacetime are called barriers if the
maximum $M$ of the mean curvature of $S_1$ is less than the minimum $m$ of 
the mean curvature of $S_2$. In that case it can be shown that $S_2$ lies to
the future of $S_1$ and that for any $H$ with $M<H<m$ there exists a CMC
hypersurface with mean curvature $H$. In particular this rules out gaps
in the set of values taken on by the mean curvature of CMC hypersurfaces
in a given spacetime. This set of values must be an interval. Apart from
the exceptional static case mentioned above, it is an open interval $I$.
The question which remains is what the endpoints of this interval are
in a given spacetime, i.e. the question of the range of the mean curvature.

It is easy to produce examples of spacetimes not containing a CMC hypersurface
by cutting pieces out of a given spacetime. For instance, one could start with
a $k=1$ Friedman model, whose homogeneous hypersurfaces define a
global CMC foliation with the mean curvature taking on all real values. 
A connected subset of this spacetime which does not contain any homogeneous
hypersurface but does admit a compact Cauchy hypersurface obviously does
not contain any compact CMC hypersurface. For any such hypersurface would
be a compact CMC hypersurface in the original spacetime and so, by
uniqueness, would have to be one of the homogeneous hypersurfaces, a
contradiction. The way to avoid such trivial examples is to consider
only maximal globally hyperbolic spacetimes, i.e. spacetimes which are
the maximal Cauchy development of initial data defined on some compact
spacelike hypersurface. The answer to the existence question in the maximal
globally hyperbolic case is not always positive. Bartnik [1] exhibited a 
cosmological spacetime which is maximal globally hyperbolic but does not admit
any compact CMC hypersurface or even any compact hypersurface where the mean 
curvature is everywhere of a single sign. 

Even if a CMC hypersurface does
exist the range of the mean curvature need not consist of all real numbers.
This follows from the following well-known argument. The Hamiltonian
constraint implies that the scalar curvature of the induced metric of
a maximal hypersurface in a spacetime satisfying the weak energy condition
must be non-negative. On the other hand, results of Gromov and Lawson [5]
imply that not all manifolds admit metrics with non-negative scalar curvature.
In fact, in a certain vague sense, most compact three-dimensional manifolds
do not do so. Each Riemannian metric on a compact three-dimensional manifold
is conformal to a metric with scalar curvature $1$, $0$ or $-1$. For each
metric only one of these three cases can occur. This is a special case of
the Yamabe theorem. According to which case occurs Riemannian metrics fall
into three classes, denoted by $Y_+$, $Y_0$ and $Y_-$ respectively. 
Three-dimensional manifolds can be classified into three disjoint types,
which are characterized by the following properties [6]. Manifolds of type 
$Y_-$
admit no metrics of non-negative scalar curvature. Manifolds of type $Y_0$
admit flat metrics and no non-flat metrics of non-negative scalar curvature.
Finally, manifolds of type $Y_+$ admit metrics of all three Yamabe classes.
The above argument shows that, provided the weak energy condition holds, a
maximal hypersurface must be a manifold of type $Y_-$ or $Y_0$. Moreover, if
it is of type $Y_0$ the second fundamental form and the energy density must
vanish on that hypersurface. In the presence of the dominant energy condition
this implies that the spacetime must be flat. Thus to produce an example of
a maximal globally hyperbolic spacetime which contains a CMC hypersurface but 
no maximal hypersurface, it suffices to do the following. Consider the maximal 
Cauchy development of data of non-zero constant mean curvature on a manifold 
of Yamabe type $Y_-$ with a matter model which satisfies the weak energy 
condition. It would also suffice to take non-vacuum data on a manifold of
type $Y_0$ and a matter model satisfying the dominant energy condition. 
A simple example of this is given by a $k=0$ Friedman model, compactified
so as to have the spatial topology of a three-dimensional torus.

Suppose now that $(M,g_{\alpha\beta})$ is a maximal globally hyperbolic
cosmological spacetime which admits at least one compact CMC hypersurface.
If $\tr k$ is non-zero on this hypersurface, it can be assumed without
loss of generality to be negative. For the sign of $\tr k$ depends on the
choice of orientation of the normal vector to the hypersurface. As a result 
of the arguments given above, if the spacetime is not flat and has a Cauchy 
hypersurface of type 
$Y_-$ or $Y_0$ and satisfies the dominant energy condition the interval $I$
of values attained by the mean curvature of compact CMC hypersurfaces must
be a subset of $(-\infty,0)$. In the case of a manifold of type $Y_+$ the
argument gives no restrictions. 

Are there restrictions coming from other sources? In fact the choice of
matter model is of importance here. A simple example which indicates the
nature of the problem is that of dust. Solutions of the Einstein equations
coupled to dust frequently develop shell-crossing singularities. Whether
a global CMC foliation exists in these spacetimes depends on whether the
foliation runs into these singularities in finite time or whether it avoids
them. Foliations by CMC hypersurfaces are known (chiefly from experience
with numerical calculations) to have a tendency to avoid spacetime 
singularities. On the other hand it is doubtful whether shell-crossing
singularities should really be considered as spacetime singularities or 
just as some mathematical pathology of the matter model. It is therefore
plausible that CMC hypersurfaces should not take these singularities
seriously and collide with them. It will be shown elsewhere that this
is in fact the case. Given $\epsilon>0$ there exist initial data for the 
Einstein equations coupled to dust with mean curvature $H<0$ such that
the maximal Cauchy development of this initial data contains no compact
CMC hypersurface of mean curvature $H'$ with $|H-H'|\ge\epsilon$. Thus
to get good general theorems it is necessary to make some detailed
assumptions on the matter model used. In the following the simplest case,
that of vacuum, will be emphasized but results for other types of matter
will be mentioned when they exist. 

\noindent
{\bf Conjecture 1} Let $(M,g_{\alpha\beta})$ be a maximal globally hyperbolic
vacuum spacetime with compact CMC Cauchy hypersurface $S$. Then if $S$ is
of type $Y_+$ the spacetime admits a foliation of compact CMC hypersurfaces
with the mean curvature taking on all real values. If $S$ is of type $Y_-$,
or of type $Y_0$ and the spacetime is non-flat, then it admits a foliation of 
compact CMC hypersurfaces taking on all values in the interval $(-\infty,0)$. 
 
\vskip 10pt
There are no known counterexamples to this conjecture. There is however one
piece of evidence which casts doubt on the form of the conjecture just given.
Numerical work of Abrahams and Evans [7] suggests that localized zero mass
singularities may occur in solutions of the Einstein vacuum equations. It
is not obvious whether singularities of this type would be avoided by CMC
foliations. More generally, it is doubtful whether all naked singularities
would be avoided. Thus there may be a close relationship between the
existence of global CMC foliations and cosmic censorship. Violations of
the simple version of cosmic censorship which says that, for reasonable
matter, no naked singularities arise in the evolution of regular initial
data, could lead to violations of Conjecture 1. To take account of this
the conjecture could be modified so as to require the conclusions only
to hold for generic initial data. 

This conjecture says nothing about which part of the spacetime is
covered by the CMC foliation. This is the subject of the next conjecture.

\noindent
{\bf Conjecture 2} If a spacetime satisfies the conditions of Conjecture
1 and the Cauchy hypersurface $S$ is of type $Y_+$ then the CMC foliation
covers the entire spacetime. If $S$ is of type $Y_-$ or $Y_0$ and admits
a spacelike Killing vector field without fixed points, then once again the 
CMC foliation covers the entire spacetime. Moreover the spacetime is 
future geodesically complete.

The first part of this conjecture is a consequence of Conjecture 1. This
is proved using the Raychaudhuri equation as in Hawking's singularity
theorem. The motivation for the other parts of the conjecture will now be 
explained. It is connected
with the idea of formation of black holes. The picture is that in the
case where $S$ is of type $Y_+$, we have a universe which recollapses.
Thus to the extent that black holes form, they will coalesce with the
cosmological singularity as the latter is approached. On the other hand,
in the case where $S$ is of type $Y_-$, the spacetime expands for ever
and black holes which form during the expansion can be expected to have 
some features in common with black holes in asymptotically flat spacetimes.
Now it is well known that in the Schwarzschild solution, for instance,
slices of non-positive mean curvature which come from outside the black hole 
cannot approach the singularity [8]. Thus such slices cannot cover the
maximally extended Schwarzschild spacetime. By analogy it seems plausible
that when black holes form in a universe which expands indefinitely, the
cosmological CMC foliation will only be able to penetrate the black hole
regions to a limited extent and so will not cover the maximal Cauchy 
development of an appropriate Cauchy hypersurface. The part of Conjecture 2 
concerned with spacetimes possessing a fixed-point free spacelike Killing
vector comes from the intuition that collapse to a black hole is necessarily
a localized phenomenon, while the presence of a spacelike Killing vector
without fixed points
forces any collapse which happens to be spread out in at least one direction.

Conjecture 2 implies the following conjecture on solutions of the constraints.

\noindent
{\bf Conjecture 3} A solution of the vacuum constraints on a compact
manifold $S$ of type $Y_-$ or $Y_0$
which admits a symmetry without fixed points does not contain
a future trapped surface $T$ with the property that the connected 
components of the inverse image of $T$ in a non-compact covering manifold 
$\tilde S$ are compact. In particular, it does not admit a future trapped
surface $T$ with spherical topology.

The proof that Conjecture 2 implies Conjecture 3 is to apply the Penrose
singularity theorem to the maximal Cauchy development of the data on
the covering space obtained by pulling back data on $S$. If
one wanted to disprove Conjecture 2 a good starting point might be to
try to construct initial data violating the conclusion of Conjecture 3.   

The idea behind Conjecture 2 is closely related to the closed universe
recollapse conjecture[9]. Define the {\it lifetime} of a cosmological
spacetime to be the supremum of the lengths of all timelike curves. This
may a priori be finite or infinite. The closed universe recollapse 
conjecture says that if the topology of the Cauchy hypersurface is of type
$Y_+$ then the lifetime is finite. Note that, in contrast to the case of
the above conjectures on CMC hypersurfaces, singularities arising as a 
result of the behaviour of the matter fields tend to help this conjecture
to be true rather than to hurt it. The case of this conjecture for
vacuum spacetimes admitting a compact CMC hypersurface is
implied by Conjecture 1, using the Raychaudhuri argument from the Hawking 
singularity theorem once more. A related conjecture going in the 
opposite direction, and which does require restrictions on the matter
fields, is that a cosmological solution of the vacuum equations, for
instance, with a Cauchy hypersurface of type $Y_0$ or $Y_-$ should always have
infinite lifetime. In the case of spacetimes which admit a CMC Cauchy 
hypersurface
this would follow from Conjecture 2. However it also makes sense in the
absence of CMC hypersurfaces. 

Note for comparison that in the case of the Einstein equations with negative
cosmological constant $\Lambda$ the above topological obstruction to the
existence of a maximal hypersurface does not occur. Thus Conjectures 1 and 2
may be modified in that case to say that for $\Lambda<0$ there is always a
global CMC foliation with the mean curvature taking all real values. Related
to this is the fact that for $\Lambda<0$ a version of the closed universe
recollapse conjecture can be proved rather easily. For this sign of the
cosmological constant the lifetime of any globally hyperbolic solution of the
Einstein-matter equations satisfying the strong energy condition
is bounded by $\pi\sqrt{-3/\Lambda}$. This statement, which follows from 
Theorem 11.9 of [10], does not even require the assumption of a compact Cauchy
hypersurface. The bound is sharp since it becomes an equality for a suitable 
globally hyperbolic part of anti-de Sitter space.  

What happens in the case that the CMC foliation does not cover the spacetime?
Looking at the Schwarzschild solution again makes it tempting to speculate
that in the case that the region covered by CMC hypersurfaces is not the
whole spacetime its boundary 
is a smooth maximal hypersurface which is non-compact. It
would consist in general of many connected components, the number of these
corresponding to the number of black holes present at late times. Each
component would have an asymptotically cylindrical geometry and would
resemble the limiting maximal hypersurface in the Schwarzschild spacetime.

The region to the future 
of the limiting maximal hypersurface in the Schwarzschild
spacetime can be covered by a CMC foliation whose leaves are non-compact,
but have complete intrinsic geometry - the situation bears some resemblance
to that in a cosmological spacetime (cf. the remarks in [8]). In the case
of the Schwarzschild solution itself the CMC hypersurfaces can be 
compactified to have topology $S^2\times S^1$, so that a cosmological
spacetime of Kantowski-Sachs type is obtained. In more general cases
(e.g. that of the Oppenheimer-Snyder solution) such a compactification
would not be possible but the evolution might nevertheless be similar to
that of a cosmological spacetime. Thus one may conjecture that in the
case of vacuum cosmological spacetimes where the region $C$ covered by
compact CMC hypersurfaces is not all of spacetime, the complement of
$C$ can be covered by a unique foliation by complete non-compact CMC
hypersurfaces.

A possible extension of Conjecture 2 is:

\noindent
{\bf Conjecture 4} If a spacetime satisfies the conditions of Conjecture
1, the Cauchy hypersurface $S$ is of type $Y_-$ or $Y_0$ and the
initial data are sufficiently close to those of a spacetime which is
covered by a foliation by compact CMC hypersurfaces then the given spacetime
also admits a global CMC foliation. Moreover the spacetime is future 
geodesically complete.

\vskip 10pt\noindent
This conjecture, if true, would be a cosmological analogue of the famous
theorem of Christodoulou and Klainerman [11] on the global nonlinear
stability of Minkowski space. 

\vskip 1cm\noindent
{\bf 2. Positive results}

The simplest case in which to investigate the truth of these conjectures
is that of spatially homogeneous spacetimes. The only classes of spatially
homogeneous spacetimes which admit a compact Cauchy hypersurface are 
those of Bianchi types I and IX and the Kantowski-Sachs spacetimes.
A much wider variety can be obtained by looking at {\it locally} spatially
homogeneous spacetimes, i.e. those whose universal covers are spatially
homogeneous. Then many other Bianchi types are possible. The specialization
of Conjectures 1 and 2 to the locally spatially homogeneous case was proved
in [12]. The manifolds obtained by compactifying spacetimes of Bianchi type
IX and Kantowski-Sachs spacetimes are of type $Y_+$. All the other Bianchi
types lead to manifolds of type $Y_0$ or $Y_-$. The result of [12] is as
follows:

\noindent
{\bf Theorem 1} Conjectures 1 and 2 hold for locally spatially homogeneous 
spacetimes, as do the analogous statements when the vacuum equations are
replaced by the Einstein equations coupled to a perfect fluid with 
reasonable equation of state or collisionless matter satisfying the
Vlasov equation.

\noindent\vskip 10pt
The most difficult case of this theorem is that of Bianchi IX spacetimes
where the proof relies on a result of Lin and Wald [13]. For inhomogeneous
cosmological spacetimes the assumption of global symmetry still only allows
a few types of group action and it is once again useful to look at local 
symmetry, where the symmetry group only acts on the universal covering
space. The simplest situation is that of a three-dimensional group acting
on two dimensional orbits. The corresponding spacetimes contain no
gravitational radiation and so it is not surprising that in vacuum they
are (locally) spatially homogeneous. For that reason the vacuum spacetimes 
in these classes are covered by Theorem 1.
For spacetimes with matter the following theorem has been proved in [14] and 
[15] under an additional restriction.

\noindent
{\bf Theorem 2} The analogues of Conjectures 1 and 2 when the vacuum 
equations are replaced by the Einstein equations coupled to collisionless 
matter satisfying the Vlasov equation or a massless scalar field hold for 
spherically symmetric spacetimes on $S^2\times S^1$. Moreover for 
spacetimes with plane or hyperbolic symmetry the analogue of Conjecture
1 holds.

\noindent\vskip 10pt
The additional restriction, namely that the mass function should be positive
on the initial hypersurface in the case of hyperbolic symmetry, can be 
removed using the techniques of [16].
Plane and hyperbolic symmetry mean that the group action on the universal
cover is an action with two dimensional orbits of the Euclidean group or
the identity component of the isometry group of the hyperbolic plane,
respectively. All these spacetimes possess three local Killing vectors.
Note that $S^2\times S^1$ belongs to the class $Y_+$, while the manifolds
occurring in the cases of plane and hyperbolic symmetry belong to the
classes $Y_0$ and $Y_-$ respectively. The conclusions of Theorem 2 for
spherical, plane and hyperbolic symmetry represent inhomogeneous
generalizations of the Kantowski-Sachs, Bianchi I and Bianchi III cases of 
Theorem 1 respectively. There is also another kind of 
spherical symmetry. This corresponds to the action of the isotropy group
of some vector on the unit sphere in $R^4$ in the group $SO(4)$ of rotations
of $R^4$. There the Killing vectors have fixed points and this makes 
the problem much more difficult. If one tries to factor out by the isometry
group, the equations on the resulting quotient space are singular. This
symmetry is closely related to the kind of spherical symmetry familiar
in the asymptotically flat case. There the equations written in polar
coordinates are singular at the centre of symmetry.

Theorems 1 and 2 together prove Conjecture 1 and its analogue
for appropriate matter models in all known cases where there are at least three
local Killing vectors and the nature of the orbits of the group action does 
not change from point to point. Under additional assumptions they also prove
Conjecture 2. The next theorem generalizes these results to certain
cases where there are only two Killing vectors[16].

\noindent
{\bf Theorem 3} Conjecture 1 holds for spacetimes with local 
$U(1)\times U(1)$ symmetry, as do the analogous statements when the vacuum 
equations are replaced by the Einstein equations coupled to collisionless 
matter satisfying the Vlasov equation or a wave map.

\noindent\vskip 10pt
The analogues of Conjectures 1 and 2 for these types of matter hold for
locally $U(1)\times U(1)$-symmetric spacetimes with cosmological constant
$\Lambda<0$.
The assumption of local $U(1)\times U(1)$ symmetry means, informally stated,
that the group action on the universal covering space looks locally like
the standard action of the group $U(1)\times U(1)$ on the torus
$S^1\times S^1\times S^1$ by rotating two of the three factors. It
includes the case of global $U(1)\times U(1)$ symmetry, where the Cauchy
hypersurface is the torus with the standard action. A wave map (also
known as a hyperbolic harmonic map or nonlinear $\sigma$-model) is a
certain kind of mapping of spacetime into a complete Riemannian manifold. In 
the case where this Riemannian manifold is the real line the equation for
wave maps reduces to the ordinary massless wave equation and so the 
matter models allowed in Theorem 3 include those covered by Theorem 2.
In fact the plane symmetric case of Theorem 2 is a special case of
Theorem 3, as are the Bianchi type II and Bianchi type VI${}_0$ cases
of Theorem 1. In should be noted that the special case of Theorem 3
for Gowdy spacetimes on $S^1\times S^1\times S^1$ was proved earlier by 
Isenberg and Moncrief[17]. The Gowdy spacetimes on the torus are those 
spacetimes with global $U(1)\times U(1)$ symmetry which satisfy the
Einstein vacuum equations and the additional condition that the so-called
twist constants vanish. There are other types of Gowdy spacetimes defined
on $S^3$ and $S^2\times S^1$ but their Killing vectors have fixed points.
Although quite a bit is known about these spacetimes ([18],[19]), Conjectures
1 and 2 remain open in that case. The next obvious step would be to look
at spacetimes with only one Killing vector, but no general results are
known in that case. 

The results which have been listed above are the most general known to the
author. It is interesting to note that the analogue of Conjectures 1 and 2
in spacetime dimension 3 have been proved by Andersson and Moncrief[20]. In 
my opinion it is not reasonable to expect that they hold in dimensions
greater than 4, at least without the additional genericity requirement.

\vskip 1cm\noindent
{\bf 3. Constant mean curvature foliations seen from the inside}

One way of thinking about the question of CMC hypersurfaces is to start
with a given spacetime and ask which part of it can be covered by a
CMC foliation. The proofs of the above theorems make use of a different
point of view, which may also be useful in more general situations. The
idea is to construct the spacetime, which is the solution of a Cauchy
problem, and a CMC foliation of it simultaneously. In other words, consider
a spacetime which is globally foliated by CMC hypersurfaces with the mean
curvature taking values in some interval. Describe the spacetime in terms
of the time coordinate defined by the mean curvature. This gives rise to 
a certain set of partial differential equations defined on some time 
interval. To prove the above theorems what is needed is to obtain a global
in time existence theorem for these equations. (The word global here refers
to a solution which is defined on the entire time interval predicted by
the conjecture under consideration.) The reason that the theorems are not 
enough
to prove the part of Conjecture 2 concerning geodesic completeness is that
in order to prove that it would be necessary to prove not only global
existence but also some facts about the qualitative behaviour of the
solution as the time parameter approaches its limiting value.

The theorem to be proved has now been reformulated as a question of global
existence. It can be reduced further to the task of obtaining sufficiently
strong bounds on the behaviour of the solution on a finite interval of
CMC time. The rough idea is as follows. Suppose a solution is given on the
interval $(t_1,t_2)$. If it is possible to obtain sufficiently strong
control on the behaviour of the solution as $t$ approaches $t_2$ (in other
words to show that the solution is not becoming singular there) then it
can be concluded that the solution extends to an interval $(t_1,t_3)$
with $t_3>t_2$.
If this control can be obtained {\it whatever the value} of the time $t_2$
then considering the longest time interval on which a solution exists
shows that global existence must hold. Thus, in order to prove global
existence it suffices to obtain strong enough estimates on the solution
on a finite time interval whose endpoint is different from that predicted
by the conjectures. To prove future geodesic completeness or to determine
in detail the nature of the initial singularity would require strong enough
estimates on a time interval which is infinite or has an endpoint where 
the solution is expected to stop existing. This is a much more difficult
task.

It will now be considered, what bounds can be obtained without any symmetry
assumptions. It follows from the CMC condition that the lapse function
$\alpha$ satisfies an elliptic equation (the lapse equation). Looking
at this equation at a point where $\alpha$ takes its maximum on a given
CMC hypersurface shows that $\alpha\le 3/t^2$. Thus an upper bound for
the lapse function away from maximal hypersurfaces is obtained. No such
bound can be hoped for near a maximal hypersurface for the following reason
(cf. [15]). In the case of those exceptional maximal hypersurfaces where
the mean curvature fails to be a good time coordinate the lapse function
blows up uniformly as the maximal slice is approached. (In the case 
$\Lambda<0$ the estimate for $\alpha$ is improved to $\alpha\le (\f13 t^2
-\Lambda)^{-1}$ and this is the source of the stronger results which can
be proved in that case.) In some highly
symmetrical cases a positive lower bound for $\alpha$ on any finite time 
interval can also be obtained. (No collapse of the lapse on a finite time
interval.) However this may well fail in general. On any finite interval
where an upper bound on $\alpha$ is available the volume of the CMC 
hypersurfaces is bounded from above and from below by a positive constant.
These are the only useful general bounds known to the author and so it
seems that there is a long way to go in order to prove the conjectures 
by this route. In the proofs of Theorems 2 and 3 the next essential step
is to use a generalization of a technique introduced by Malec and 
\'O Murchadha[21] and no way is known to generalize this method to
spacetimes with less than 2 local Killing vectors.

To conclude, some remarks on the relationship of the conjectures above to
cosmic censorship will be made. The idea of using a preferred time coordinate
to reformulate the (strong) cosmic censorship hypothesis as a statement
about global existence (and asymptotic behaviour) of solutions of a system
of partial differential equations was put forward by Eardley and Moncrief[3].
Here the first step is to prove global existence. However, to prove something
about strong cosmic censorship the asymptotic behaviour of the solution
must be controlled. To make this idea concrete, consider the relationship
of Theorems 1, 2 and 3 to cosmic censorship in the respective classes of
spacetimes. In the situation of Theorem 1 it can be shown for non-vacuum
spacetimes that the initial singularity is a curvature singularity, so
that the specialization of strong cosmic censorship holds. However the
asymptotic behaviour of the solution near the singularity is poorly
understood, even in the Bianchi type I case [22]. In the situation of 
Theorem 2, it was shown by Rein [23] that the initial singularity is
a curvature singularity for an open set of initial data (but not for all
initial data). In the case of Gowdy spacetimes a similar result was
proved by Chrusciel [24] while in the more special case of the polarized
Gowdy spacetimes, detailed information can be obtained without additional
restrictions on the initial data[19]. Clearly a lot remains to be done in
this direction. Moreover, none of the results on inhomogeneous spacetimes
just mentioned was proved using a CMC time coordinate. Instead a time
coordinate was used 
which was specially adapted to the symmetry of the models. It would be
desirable to prove versions of these results using CMC slicing in the hope
of unifying, strengthening and generalizing them.

\noindent
{\bf References}

\noindent
[1] Bartnik, R.: Remarks on cosmological spacetimes and constant mean
curvature hypersurfaces. {\it Commun. Math. Phys.} 117 (1988) 615-624.
\next
[2] Marsden, J. E., Tipler, F. J.: Maximal hypersurfaces and foliations of
constant mean curvature in general relativity. {\it Phys. Rep.} 66 (1980)
109-139.
\next
[3] Eardley, D., Moncrief, V.: The global existence problem and cosmic
censorship in general relativity. {\it Gen. Rel. Grav.} 13 (1981) 887-892.
\next
[4] Beig, R.: The classical theory of canonical general relativity. In:
J. Ehlers, H. Friedrich: Canonical gravity: from classical to quantum.
Springer, Berlin, 1994.
\next
[5] Gromov, M., Lawson, H. B.: Positive scalar curvature and the Dirac 
operator on complete Riemannian manifolds. {\it Publ. Math. IHES} 58 (1983)
295-408.
\next
[6] Besse, A.: Einstein manifolds. Springer, Berlin, 1987.
\next
[7] Abrahams, A. M., Evans, C. R.: Universality in axisymmetric vacuum
collapse. {\it Phys. Rev.} D70 (1994) 3998-4003.
\next
[8] Eardley, D. M., Smarr, L.: Time functions in numerical relativity: 
marginally bound dust collapse. {\it Phys. Rev.} D19 (1979) 2239-2259.
\next
[9] Barrow, J. D., Tipler, F. J.: The closed-universe recollapse conjecture.
{\it Mon. Not. R. Astron. Soc.} 223 (1986) 835-844.
\next
[10] Beem, J. K., Ehrlich, P. E. and Easley, K. L.: Global Lorentzian
geometry (2nd Ed.) Marcel Dekker, New York, 1996.
\next
[11] Christodoulou, D, Klainerman, S.: The global nonlinear stability
of the Minkowski space. Princeton University Press, Princeton. 1993.
\next
[12] Rendall, A. D.: Global properties of locally spatially homogeneous 
cosmological models with matter. {\it Math. Proc. Camb. Phil. Soc.} 118 
(1995) 511-526.
\next
[13] Lin, X-F., Wald, R.: Proof of the closed universe recollapse conjecture 
for general Bianchi type IX cosmologies. {\it Phys. Rev.} D41 (1990) 
2444-2448.  
\next
[14] Rendall, A. D.: Crushing singularities in spacetimes with spherical, 
plane and hyperbolic symmetry. {\it Class. Quantum Grav.} 12 (1995) 1517-1533.
\next
[15] Burnett, G. A., Rendall, A. D.: Existence of maximal hypersurfaces in 
some spherically symmetric spacetimes. {\it Class. Quantum Grav.} 13 (1996) 
111-123.
\next
[16] Rendall, A. D.: Existence of constant mean curvature foliations in 
spacetimes with two-dimensional local symmetry. Preprint gr-qc/9505022.
\next
[17] Isenberg, J., Moncrief, V.: The existence of constant mean curvature
foliations of Gowdy 3-torus spacetimes. {\it Commun. Math. Phys.} 86 (1983)
485-493.
\next
[18] Chru\'sciel, P.: On spacetimes with $U(1)\times U(1)$ symmetric
compact Cauchy surfaces. {\it Ann. Phys.} 202 (1990) 100-150.
\next
[19] Chru\'sciel, P, Isenberg, J. and Moncrief V.: Strong cosmic
censorship in polarised Gowdy spacetimes. {\it Class. Quantum Grav.} 7 (1990)
1671-1680.
\next
[20] Andersson, L., Moncrief, V.: On the global evolution problem in 2+1
gravitation. Preprint. 
\next
[21] Malec, E., \'O Murchadha, N.: Optical scalars and singularity
avoidance in spherical spacetimes. {\it Phys. Rev.} D50 (1994) 6033-6036.
\next
[22] Rendall, A. D.: The initial singularity in solutions of the 
Einstein-Vlasov system of Bianchi type I. {\it J. Math. Phys.} 37 (1996)
438-451.
\next
[23] Rein, G.: Cosmological solutions of the Vlasov-Einstein system with
spherical, plane and hyperbolic symmetry. {\it Math. Proc. Camb. Phil.
Soc.} 119 (1996) 739-762.
\next
[24] Chru\'sciel, P. T.:  On the uniqueness in the large of solutions of
Einstein's equations. (Strong cosmic censorship.) Proc. Centre Math. Anal.,
Australian National University, Vol 27, 1991.

\end